# Spectral characterization of long-lived luminescence in h-BN nanopowder under UV excitation

Ilya Weinstein[1,2*], Dmitriy Spiridonov[1], Alexander Vokhmintsev[1], Ahmed Henaish [1,3]

1   NANOTECH Centre, Ural Federal University, 620002 Ekaterinburg, Russia
2   Institute of Metallurgy of the Ural Branch of the Russian Academy of Sciences, 620016 Ekaterinburg, Russia
3   Physics Department, Faculty of Science, Tanta University, Tanta, Egypt
*   Correspondence: i.a.weinstein@urfu.ru; Tel.: +7 343 375 93 74

**Abstract:** Photoluminescence (PL) features of nanostructured h-BN powder are studied in the range of 200 – 600 nm in millisecond time window. It is found that four PL excitation bands of 4.58, 5.01, 5.29, and 5.77 eV are characteristic of the spectral region at hand. It is shown that the PL emission spectra can be quantitatively described through a superposition of three Gaussian-shaped peaks: 2.6, 3.1, and 3.7 eV. The observable luminescence is established to be due to recombination processes involving centers whose energy levels in the bandgap are formed by impurity oxygen $O_N$ and carbon $C_N$, as well as nitrogen vacancies $V_N$ and related complexes.

## 1. Introduction

Micro- and nanosized hexagonal boron nitride (h-BN) structures are regarded as the basis for creating of a promising highly sensitive media for new photonic and optoelectronic applications [1–3]. A wide bandgap $E_g$ > 5.2 eV [4, 5 and refs. in it, 6], laser generation detected in the ultraviolet range [7] enable one to judge h-BN as of an efficient emitter and detector material in the UV and visible radiation regions [8–10]. It is known [11–16] that oxygen and carbon are the main and, as a rule, uncontrolled technological impurities in nitrides. The electrophysical and luminescent properties of h-BN powders are due to a system of electron and hole traps formed by similar defects and complexes based on them [11, 17–21].

Currently great research interest is attracting to the UV fluorescence in h-BN structures in the range of 215 – 400 nm [2, 14, 19]. Radiative short-wavelength relaxation, as a rule, is associated with the free and bound excitonic states decay [2, 22]. Recombination luminescence at λ > 300 nm is caused by intrinsic and impurity defects, including donor – acceptor pairs (DAP) based on carbon – oxygen centers [11, 12, 14, 19, 23]. Earlier [17, 22, 24] it was shown using time-resolved spectroscopy that short-lived luminescence in the UV region was effectively excited by band-to-band and subband transitions and had different decay kinetics in two time gates: a fast one with 1 – 4 ns and a slow one with 22 – 200 ns. Under band-to-band excitation the kinetic dependences are multiexponential, the contribution of slow processes to the averaged lifetime of electronic excitations increases due to the arising separation of charge carriers and the presence of trap states in the h-BN energy gap [17, 19]. In turn, the mechanisms of long-lived millisecond luminescence in the UV and visible ranges, which is observed mostly in thermally stimulated processes [11, 18, 20, 25], affect significantly the spectral characteristics of h-BN at room temperature and above. In this case, when developing optoelectronic elements based on h-BN structures and optimizing their applied parameters, it is necessary to take into account the peculiarities of the slow emission and to understand the basic regularities of recombination mechanisms involving separated charges also. The present work investigates the photoluminescence properties in millisecond time window of a nanostructured h-BN powder with low carbon and oxygen impurity content.

## 2. Materials and Methods

The research was conducted for h-BN powder produced by Hongwu International Group Ltd., Hong Kong. According to the manufacturer, the powder contains carbon and oxygen impurities, with their total concentration not exceeding 0.5 wt %. An analysis of SEM images [6], showed that the powder's particle size varies in the range from 40 to 1100 nm and is well described by a lognormal distribution with a 300 nm median and a 226 nm mode. Based on the data of [6], it can be argued that the studied structure of h-BN is mesographitic with a graphitization index GI = 9.

Photoluminescence measurements were taken at room temperature using a Perkin Elmer LS 55 spectrometer in the regime of long-lived processes: emission recording began with a delay of 0.05 ms after an excitation pulse; data collection time and the duration of one cycle were 12.5 ms and 20 ms, respectively. The excitation of photoluminescence was brought about in the range of 200-280 nm with a step of 1 nm. The emission intensity (I) was recorded by scanning the spectral region of 290-600 nm at a speed of 120 nm/min. The slit width of the excitation and emission channels amounted 10 nm and 20 nm, respectively. For numerical analysis, the $I(\lambda)$ spectral dependencies measured were translated into the $I(E)$ energy spectra, accounting for the correction by $\lambda^2$.

## 3. Results

The experimental PL spectra in the coordinates "intensity I – excitation wavelength $\lambda_{exc}$ – emission wavelength $\lambda_{em}$" are shown in Figure 1a. Figure 1b outlines the top view projection of the 3D plot specified, plotted in the form of a luminescence intensity mapping. It can be seen that the emission dependencies have several distinct maxima at 415 and 345 nm. The most intense peak is detected at $(\lambda_{em}, \lambda_{exc}) \approx (415, 215)$ nm. The same band includes less pronounced maxima (415, 234), (416, 250), and (410, 273) nm, see Fig. 1b. The luminescence in the $\lambda_{em} \approx 345$ nm is characterized by excitation in the UV region with maxima of $\lambda_{exc} \approx 218$, 236, and 250 nm.

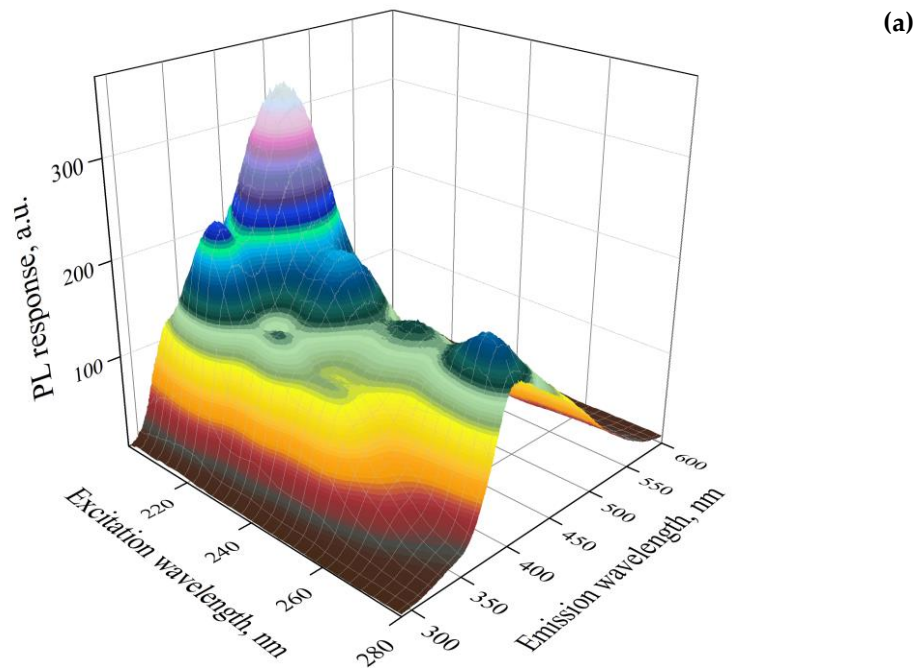

(a)

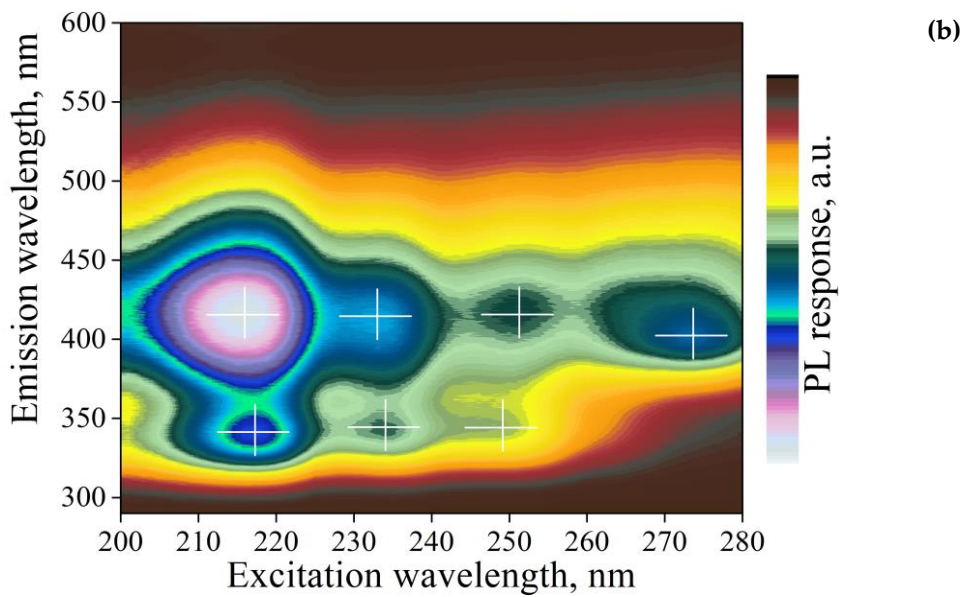

**Figure 1.** Photoluminescence in h-BN nanopowder: (**a**) 3D plot; (**b**) top view projection.

## 4. Discussion

Figure 2 displays PL excitation spectra in the emission bands of $E_{em} \approx 3.6$ eV (345 nm) and 3.0 eV (415 nm). Presented experimental curves were numerically fitted by a superposition of several Gaussian components ($R^2 \geq 0.998$). The approximation results are listed in Table 1. It is seen that the component with an energy $E_I = 5.75 - 5.80$ eV and a half-width $\omega_I = 0.57 - 0.66$ eV is dominant. Recently [6] it was evaluated the bandgap for the powder under study being $E_g = 5.41$ eV. The observed excitation band is caused by band-to-band transitions with allowing for this fact.

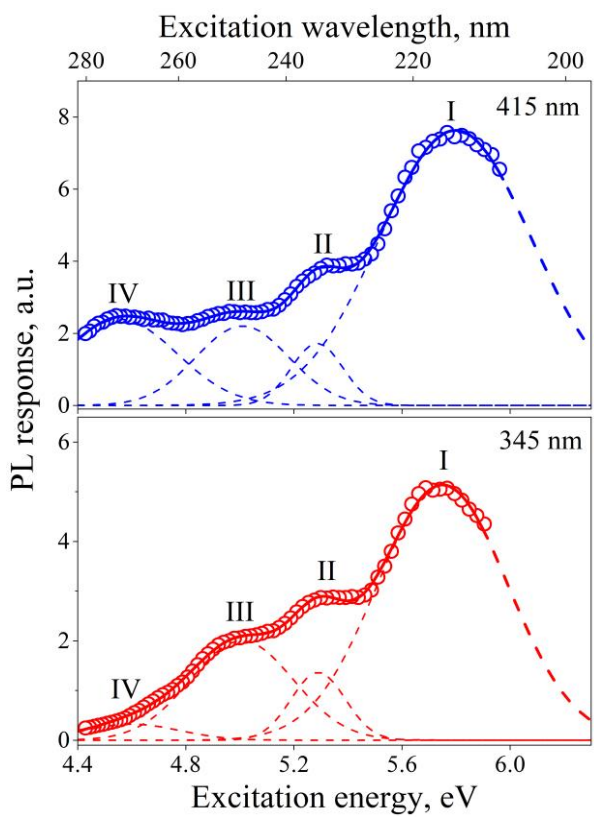

**Figure 2.** PL excitation spectra in 415 nm (**a**) and 345 nm (**b**) bands. Symbols – experimental results; dashed lines – Gaussian-shaped components; solid line – resulting curve.

**Table 1.** Results of PL excitation spectra approximation.

| Gaussian | Spectral parameter | Emission band | | Optical transitions |
|---|---|---|---|---|
| | | 3.6 eV (345 nm) | 3.0 eV (415 nm) | |
| I | $E_I$, ±0.02 eV | 5.75 | 5.80 | band-to-band VB → CB |
| | $\omega_I$, ±0.05 eV | 0.57 | 0.66 | [4, 6] |
| II | $E_{II}$, ±0.03 eV | 5.29 | 5.29 | q-DAP |
| | $\omega_{II}$, ±0.05 eV | 0.24 | 0.22 | [17, 26] |
| III | $E_{III}$, ±0.02 eV | 5.01 | 5.01 | VB → $V_N$ |
| | $\omega_{III}$, ±0.05 eV | 0.45 | 0.41 | [28] |
| IV | $E_{IV}$, ±0.03 eV | 4.60 | 4.56 | $C_N$ → CB |
| | $\omega_{IV}$, ±0.05 eV | 0.41 | 0.49 | [19, 28, 31] |

The possible origin of other excitation bands was also analyzed in the framework of known literature data. It is shown in [27, 28] that the levels formed by impurity oxygen are ionized when excited at 4.5 eV (consistent with $E_{IV}$). In this case, the electrons move to the conduction band ($C_N$ → CB). The $E_{III}$ band appears to be due to transitions of charge carriers from the valence band to the capture levels formed by nitrogen vacancies $V_N$ and complexes based on them [28]. The component $E_{II}$ = 5.29 eV is in good agreement with the results of [26], where excitation in this region was attributed to quasi-donor-acceptor pairs (q-DAP) arising from electrostatic band fluctuations induced by charged defects.

Figure 3 shows PL emission spectra upon excitation in the $E_I$ - $E_{IV}$ bands. It is clear that the dependencies measured in the range from 4.77 to 6.20 eV have a similar shape: two peaks close in intensity are observed at 3.65 and 3.02 eV. When exciting the powder by photons with an energy E < 4.77 eV, the luminescence drops in the region of 3.65 eV. As a result, a peak of 3.1 eV dominates.

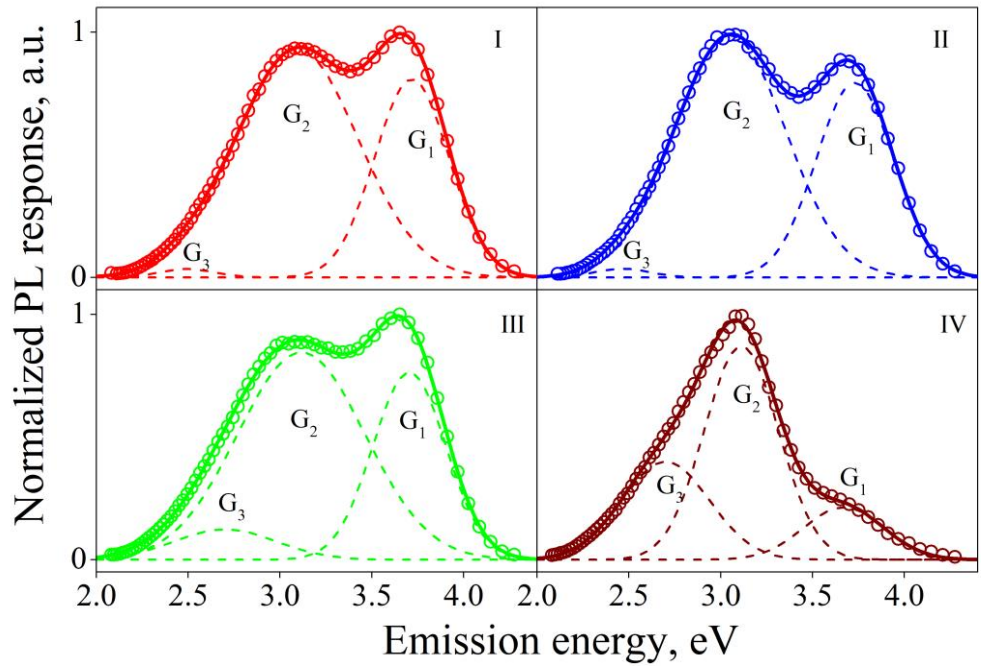

**Figure 3.** Emission spectra upon excitation in 5.77 eV (I), 5.30 eV (II), 4.96 eV (III) и 4.59 eV (IV) bands and their approximation. Symbols – experimental results; dashed lines – Gaussian-shaped components; solid line – resulting curve.

All recorded emission spectra were deconvoluted into three Gaussian-shaped components with maxima $E_1$ = 3.70 ± 0.04 eV, $E_2$ = 3.10 ± 0.07 eV, and $E_3$ = 2.6 ± 0.1 eV (the latter being weak enough). The obtained curves and calculated parameters are given in Figure 3 and Table 2, respectively. It should be underscored that these data align well with the estimates for h-BN micropowders synthesized by the plasma-enhanced CVD method [25]. This method yields emission maximum bands at 3.61, 3.27, and 2.62 eV. The foregoing indicates related mechanisms and similar structural defects that are responsible for the photoluminescence explored.

**Table 2.** Results of PL emission spectra approximation.

| Gaussian | Spectral parameter | Excitation band I | II | III | IV | Responsible defects |
|---|---|---|---|---|---|---|
| $G_1$ | $E_1$, ± 0.02 eV | 3.72 | 3.73 | 3.70 | 3.67 | $C_N$, $O_N$ [12] |
| | $\omega_1$, ± 0.04 eV | 0.48 | 0.50 | 0.47 | 0.51 | $V_N$, $O_N$ [14] surface $V_N$ [29] |
| $G_2$ | $E_2$, ± 0.07 eV | 3.09 | 3.05 | 3.12 | 3.11 | one-boron center, $C_N$ [11, 30] |
| | $\omega_2$, ± 0.05 eV | 0.78 | 0.71 | 0.82 | 0.48 | $V_N$, $O_N$ [13] $V_N$, 2-$V_N$, 3-$V_N$ [29] |
| $G_3$ | $E_3$, ± 0.05 eV | 2.50 | 2.47 | 2.70 | 2.70 | $C_N$ [23, 31] |
| | $\omega_3$, ± 0.05 eV | 0.36 | 0.35 | 0.65 | 0.56 | $C_B$ [31] |

It is worth emphasizing that the luminescence in the $E_1$ band can be caused by donor-acceptor recombination processes involving oxygen and carbon impurities in the positions of nitrogen, $O_N$- and $C_N$-centers, respectively [12, 14]. Besides, the $E_2$ luminescence can arise due to recombination processes involving one-boron (1B-) and carbon $C_N$-centers [11, 30]. It is known [11] that 1B-centers are formed in the presence of oxygen atoms, which is consistent with the impurity composition of the samples tested. According to the results of [13, 29], single $V_N$ nitrogen vacancies, as well as their more complicated complexes of the 2-$V_N$ and 3-$V_N$ type, can provoke the luminescence in the $E_2$ region. The paper [23] reports that the $E_3$ component takes place conditional upon transitioning charge carriers between the $C_N$ level and the valence band. Besides the $G_3$ emission is possibly caused by radiative $C_B \to C_N$ transitions [31].

## 5. Conclusions

The spectral features of long-lived photoluminescence in nanostructured h-BN powder were investigated in the work. It is found that the excitation spectra have a too involved structure. In the 6.2 - 4.1 eV (200 - 300 nm) spectral region studied, four components of 5.77, 5.29, 5.01, and 4.58 eV are distinguished. Based on the earlier obtained data on the powder's bandgap width, a conclusion was made that the 5.77 eV band is due to band-to-band transitions. A comparative analysis of the experimental spectral parameters with independent literature data proved that the observable emission in h-BN nanostructured powder are mainly governed by recombination processes involving optically active centers whose energy levels are related to oxygen $O_N$ and carbon $C_N$ impurities, as well as nitrogen vacancies $V_N$ and complexes based on them.

**Author Contributions:** Conceptualization, I.W.; formal analysis, D.S.; photoluminescence investigation, A.V., A.H. and D.S.; writing—original draft preparation, A.H. and D.S.; writing—review and editing, I.W. and A.V.; visualization, D.S.; supervision, I.W.; funding acquisition, I.W. and D.S. All authors have read and agreed to the published version of the manuscript.

**Funding:** This research was supported by Russian Foundation for Basic Research (grant number 18-32-00550) and by Act 211 Government of the Russian Federation (contract № 02.A03.21.0006). A.V. and I.W. thank Minobrnauki research project for support.

**Conflicts of Interest:** The authors declare no conflict of interest.